\documentclass[conference]{IEEEtran}
\IEEEoverridecommandlockouts
\usepackage{cite}
\usepackage{siunitx}
\usepackage{amsmath,amssymb,amsfonts}
\usepackage{algorithmic}
\usepackage{graphicx}
\usepackage{textcomp}
\usepackage{xcolor}
\usepackage{changepage}
\usepackage[inline]{enumitem}
\usepackage{booktabs,multirow}
\usepackage[colorlinks]{hyperref} 
\hypersetup{
  colorlinks = true,
  linkcolor=blue,
  citecolor=blue,
  urlcolor=blue
}

\usepackage[caption=false]{subfig}
\def\BibTeX{{\rm B\kern-.05em{\sc i\kern-.025em b}\kern-.08em
    T\kern-.1667em\lower.7ex\hbox{E}\kern-.125emX}}
    
\usepackage{placeins}

\begin{document}
\title{Array Placement in Distributed Massive MIMO for Power Saving considering Radiation Pattern
\thanks{Yi-Hang Zhu is supported by the Chinese Scholarship Council (CSC). This work has been published by the 2021 IEEE 94th Vehicular Technology Conference: VTC2021-Fall. \href{http://dx.doi.org/10.1109/VTC2021-Fall52928.2021.9625546}{DOI: 10.1109/VTC2021-Fall52928.2021.9625546
}}
}

\author{\IEEEauthorblockN{Yi-Hang Zhu, Laura Monteyne, Gilles Callebaut, Fran\c{c}ois Rottenberg, Liesbet Van der Perre}
\IEEEauthorblockA{\textit{KU Leuven Department of Electrical Engineering (ESAT) DRAMCO} \\
9000 Ghent, Belgium \\
yihang.zhu@kuleuven.be}

}

\maketitle

\begin{abstract}
A distributed antenna system (DAS) consists of several interconnected access points (APs) which are distributed over an area.
Each AP has an antenna array. 
In previous studies, the DAS has been demonstrated great potential to improve capacity and power efficiency compared to a centralized antenna system (CAS) which has all the antennas located in one place.
The existing research also has shown that the placement of the APs is essential for the performance of the DAS.
However, most research on AP placement does not take into account realistic constraints.
For instance, they assume that APs can be placed at any location in a region or the array radiation pattern of each AP is isotropic.  
This paper focuses on optimizing the AP placement for the DAS with massive MIMO (D-mMIMO) in order to reduce the transmit power.
A square topology for the AP placement is applied, which is reasonable for deploying the D-mMIMO in urban areas while also offering theoretically interesting insights.
We investigate the impact of the radiation pattern, signal coherence, and region size on the D-mMIMO's performance.
Our results suggest that 
\begin{enumerate*}[label={(\roman*)}]
    \item among the three factors, the array radiation pattern of each AP is the most important one in determining the optimal AP placement for the D-mMIMO;
    \item the performance of the D-mMIMO is highly impacted by the placement and array radiation pattern of each AP, the D-mMIMO with unoptimized placement may perform even worse than the CAS with massive MIMO (C-mMIMO);
    \item with the consideration of patch antennas and the mutual coupling effect, the optimized D-mMIMO can potentially save more than \SI{7}{\decibel} transmit power compared to the C-mMIMO.
\end{enumerate*}
Furthermore, our analytical results also provide an intuition for determining an adequate AP placement for the D-mMIMO in practice.  
\end{abstract}


\section{Introduction}
A distributed antenna system (DAS) comprises interconnected antenna arrays, called access points (APs), geographically distributed over an area.
It is different from a conventional centralized antenna system (CAS), which has all the antennas located in one place.
The history of the DAS goes back to decades ago when the DAS was introduced for indoor wireless networks~\cite{saleh1987distributed}. 
Since then, the DAS has attracted a significant amount of attention due to its capacity and power efficiency advantages compared to the CAS~\cite{clark2001distributed,roh2002outage,dai2011comparative}.
In recent years, in the context of massive multiple-input multiple-output (MIMO) systems~\cite{bjornson2017massive}, the performance of specifically DAS with massive MIMO or distributed massive MIMO (D-mMIMO) has been enhanced even further~\cite{ngo2017cell}.  
 
Meanwhile, existing studies have shown that the performance of the DAS can be improved by optimizing the AP placement (i.e., the locations for deploying the APs)~\cite{shen2007optimal,gan2007sum,park2012antenna,yang2015performance,kamga2016spectral,wang2009antenna,han2010design,firouzabadi2011optimal,minasian2018rrh,zhang2020optimal}.   
Many researchers such as~\cite{gan2007sum,han2010design,park2012antenna,yang2015performance,kamga2016spectral} consider a circular topology for the APs (i.e., the APs are deployed evenly on a circle) and try to find an optimal radius for the circle. 
Other studies, such as~\cite{wang2009antenna,firouzabadi2011optimal,minasian2018rrh,zhang2020optimal}, suggest that APs should be deployed without a restricted topology and propose different algorithms for optimizing AP placements.
Those algorithms allow APs to be placed at any location in the region.
However, for practical scenarios, e.g., Manhattan-style street grid, it is typically not feasible to place the APs circularly or arbitrarily as considered in the literature.  
In those scenarios, for the sake of aesthetics and deployment simplicity, APs can be deployed along walls in the streets and be less visible \cite{van2020radioweaves}.
Placing APs on walls also ensures that the APs are less impacted by bad weather, for example, rain or wind. 
In other words, a rectangular topology of AP placement is an attractive proposal and a more realistic assumption for those scenarios. 

Moreover, since each AP has an antenna array, the mutual coupling effect between the array elements will heavily impact the power efficiency and radiation pattern of each array~\cite{chen2017finite}.
Due to coupling, the radiation pattern of the array becomes more directional even if, individually, the elements, for example, dipoles, have a little directive pattern. 
However, in the DAS-related literature, it is common to assume that each AP has an isotropic or omnidirectional radiation pattern. 
To our best knowledge, the consideration of either mutual coupling or realistic element radiation patterns in the context of DAS placement is missing.

\begin{figure}[!htb]
  \centering
  \includegraphics[width=0.4\textwidth]{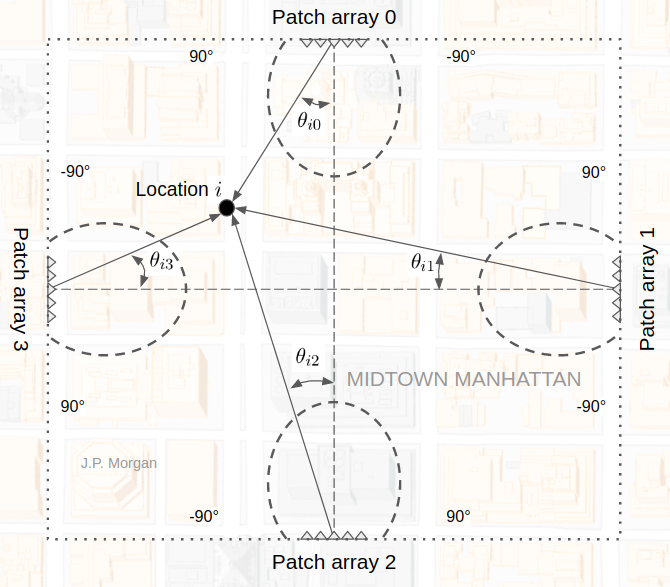}
  \caption{The coverage region (dotted) and square topology for the D-mMIMO, where the dashed curves represent the array patterns.}
  \label{fig:region}
\end{figure}

According to~\cite{chen2017finite}, an antenna array that consists of directional elements, for example, patch antennas, has less mutual coupling effect compared to one that comprises more omnidirectional elements, for example, dipoles. 
This paper considers a D-mMIMO network with each AP having an array of patch antennas.
The topology for the AP placement is considered to be square as a particular case of the rectangular one. This will only affect the algorithm that is used to solve the problem. The extension to generic rectangular regions is straightforward and not included in this paper for simplicity.
The network configuration is illustrated in Fig.~\ref{fig:region}.
We optimize the AP placement to minimize the total transmit power while guaranteeing a minimum received signal power at each possible user location.

This paper contributes to the literature three aspects.
\begin{enumerate*}[label={(\roman*)}]
  \item We study and formalize the optimization problem for the AP placement for the D-mMIMO in an urban area scenario considering a square topology for the first time.
  \item We investigate the impact of different factors on the optimal AP placement and the total transmit power. 
  The factors include mutual coupling, coverage region size, and signal coherence. 
  \item With the consideration of radiation patterns, we analyze the power saving of the D-mMIMO compared to the CAS with Massive MIMO (C-mMIMO). 
\end{enumerate*}

The rest of this paper is organized as follows. Section~\ref{sec: model} describes the problem, the system model, and the D-mMIMO network configuration. Section~\ref{sec:solver} introduces a heuristic algorithm for optimizing the AP placement. Section~\ref{sec:results} presents the analytical results and Section~\ref{sec:conclusion} concludes the paper.

\section{System Model and problem definition}\label{sec: model}

As illustrated in Fig.~\ref{fig:region}, we study the case of a D-mMIMO network with four distributed antenna arrays ($t=0,1,2,3$) in a square region. Each array is a half-wavelength spaced uniform rectangular array and positioned at an edge of the region. The arrays have $M$ elements, which we consider to be patch antennas such that the arrays could fit well on walls of buildings. 
We consider that the individual users have a single isotropic antenna. 
The black dot in Fig.~\ref{fig:region} represents one of the possible user locations which uniformly cover the region. 
The space between the locations considered in the optimization algorithm is one meter.   
Only large-scale fading, considering different path loss exponents, is incorporated to make it possible to define the mathematical optimization problem and draw statistically relevant conclusions. 
The transmit power per AP $P_{\mathtt{T}}$ is assumed to be the same for all APs.
{\it The objective of the problem is to minimize $P_{\mathtt{T}}$ by optimizing the coordinates of each AP $t$, which is denoted by variable $\mathbf{c}_{{\mathtt{T}}_{t}}$.}   


Assuming that a single-user is served per time/frequency resource, no user-interference is present.
When the signals from different APs are adding up coherently at the user location, constraint~\eqref{eq:coherent} is applied to ensure each location $i$ is covered by sufficient signal power regarding the minimum required amount $P_{\mathtt{R}}$. 
\begin{equation}
  \label{eq:coherent}
  \left(\sum_{t=0}^{3} \sqrt{P_{\mathtt{T}}G_{\theta_{it}} \frac{\lambda^{2}}{16\pi^{2}r_{it}^{n}}}\right)^{2} \ge P_{\mathtt{R}} \quad \forall i
\end{equation}

In the case where the signals from different APs are not adding up coherently, constraint~\eqref{eq:non_coherent} is applied for ensuring the signal coverage.
\begin{equation}
  \label{eq:non_coherent}
  \sum_{t=0}^{3} P_{\mathtt{T}}G_{\theta_{it}} \frac{\lambda^{2}}{16\pi^{2}r_{it}^{n}} \ge P_{\mathtt{R}}
  \quad \forall i
\end{equation}

Variable $\theta_{it}$ in the constraints is the angular direction of location~$i$ regarding the broadside of AP~$t$.
Variable $r_{it}$ is larger or equal to the far field distance of the AP $r_{0}$ and calculated via Eq.~\eqref{eq:rit}, where parameter $\mathbf{c}_{R_{i}}$ represents the coordinates of location~$i$. 
\begin{equation}
    \label{eq:rit}
    r_{it} = \max\{r_{0}, ||\mathbf{c}_{R_{i}}-\mathbf{c}_{\mathtt{T}_{t}}||_{2}\}  
\end{equation}

Parameters $\lambda$ and $n$ are the wavelength for the considered radio wave and path-loss exponent, respectively. 
The array gain of each AP $G_{\theta}$ is calculated via Eq.~\eqref{eq:array_gain}~\cite{mailloux2017phased}, where $\widetilde{G}_{\mathtt{E}_{\theta}}$ is the embedded element gain and $\theta_{0}$ is the scan angle. 
If mutual coupling is not considered, $\widetilde{G}_{\mathtt{E}_{\theta}}$ equals the gain of an isolated patch antenna $G_{\mathtt{E}_{\theta}}$.
Otherwise, given that the scan loss due to mutual coupling can be approximated as $(\cos\theta)^{\frac{3}{2}}$ for a patch array~\cite{mailloux2017phased}, 
$\widetilde{G}_{\mathtt{E}_{\theta}}$ is calculated via Eq.~\eqref{eq:element_gain}.  
This paper assumes that the transmitters have perfect channel state information and apply maximum ratio transmission at each AP.
Therefore, the array gain $G_{\theta}$ equals $M$ times of the embedded element gain $\widetilde{G}_{\mathtt{E}_{\theta}}$.

\begin{equation}
  \label{eq:array_gain}
  G_{\theta} = \frac{|\sum_{m}^{M-1} \sqrt{\widetilde{G}_{\mathtt{E}_{\theta}}}e^{\jmath\pi m (\sin\theta_{0}-\sin\theta)}|^{2}}{M}
\end{equation}

\begin{equation}
  \label{eq:element_gain}
  \widetilde{G}_{\mathtt{E}_{\theta}} =  G_{\mathtt{E}_{\theta}}(\cos\theta)^{\frac{3}{2}} 
\end{equation}



\section{Algorithm for Optimizing the AP Placement 
}\label{sec:solver}
\begin{figure}[!htb]
  \centering
  \includegraphics[width=0.2\textwidth]{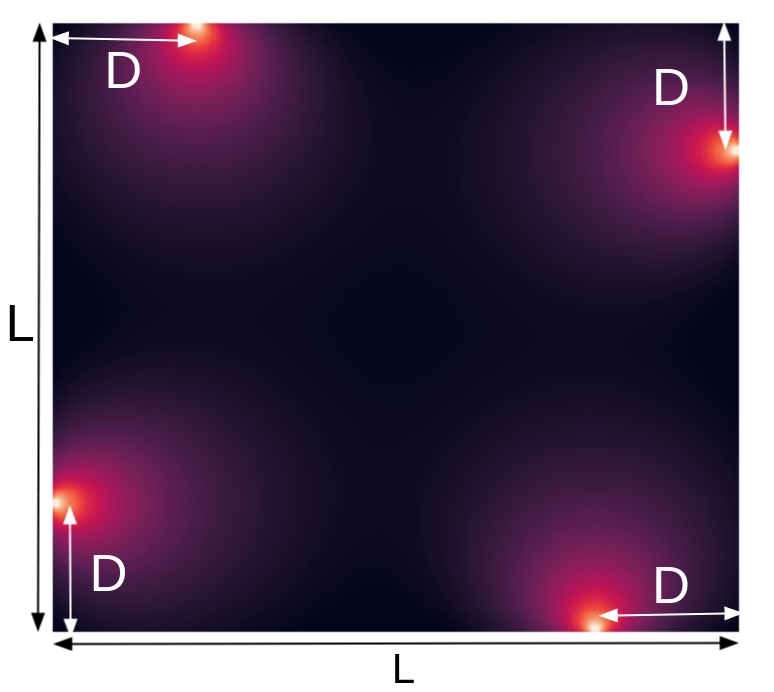}
  \caption{Example of a symmetrical AP placement for the D-mMIMO in a $L\times L$ square region, each AP is deployed at a distance of $D$ from its nearest corner.
  }
  \label{fig:placement}
\end{figure}

As has been demonstrated by \cite{gan2007sum} and \cite{han2010design} and also confirmed by our preliminary results, an optimal AP placement that minimizes the transmit power is symmetrical with respect to the region center due to the symmetry of the considered problem. 
Fig.~\ref{fig:placement} illustrates an example for such a symmetrical AP placement.


Based on the above observation, the solution space for the problem is significantly narrowed down and the problem can be solved by using a simple line search-based algorithm.
The algorithm generates an initial solution by placing the four APs at the four corners of the region. 
Then, the APs are shifting in a clockwise direction at a certain resolution.
The smaller the resolution the higher precision of the placements.
The transmit power $P_{\mathtt{T}}$ is calculated for the new placement after each shifting.
The algorithm stops after the total shifting distance is larger or equal to a criterion and the best-found placement is returned.

\section{Results and discussion}\label{sec:results}
This section analyzes the performance of the D-mMIMO with respect to the AP placement and total transmit power. 
The parameters are set as follows.
The gain pattern for a single patch antenna $G_{\mathtt{E}_{\theta}}$ is from~\cite{callebautAsilomar}. 
The number of array elements in each AP ($M$) is 32 (i.e., $4\times8$). 
The considered radio frequency is \SI{2.6}{\GHz} and the wavelength $\lambda$ is \SI{0.12}{\metre} accordingly.
$r_{0}$ is set to \SI{3}{\metre} regarding the array configuration for each AP and the Fraunhofer distance. 
According to Eqs.~\eqref{eq:coherent} and~\eqref{eq:non_coherent}, the value for the minimum received signal power $P_{\mathtt{R}}$ will only impact the transmit power per AP $P_{\mathtt{T}}$. 
It will not impact the optimal AP placement or the difference in $P_{\mathtt{T}}$ when comparing different solutions, for example, the optimal solutions for the situations with and without considering mutual coupling. 
Therefore, $P_{\mathtt{R}}$ is arbitrarily set to \SI{-60}{dBm}, which is not completely unrealistic from a practical point of view.

\begin{figure}[!htb]
    \centering
    \includegraphics[width=0.5\linewidth]{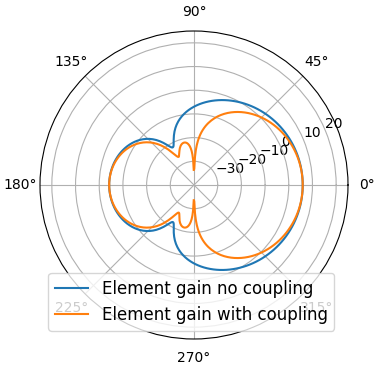}
    \caption{The element gain patterns (dBi) for a 32-element uniform rectangular patch array with and without considering mutual coupling}\label{fig:array_gain}  
\end{figure}

Fig.~\ref{fig:array_gain} illustrates the element gain 
$\widetilde{G}_{\mathtt{E}_{\theta}}$ with and without considering the mutual coupling effect. 
As shown in Fig.~\ref{fig:array_gain}, with the consideration of mutual coupling, element gain is decreasing when scanning to endfire and becomes zero at endfire. 
This increases the directivity of the array radiation pattern of each AP. 
As indicated by~\cite{mailloux2017phased}, the scan loss approximation applied in Eq.~\eqref{eq:element_gain} is for the scan angle up to \ang{60} from broadside. 
However, we need to consider the scan angle up to the endfire for each array. 
With this scan loss approximation, the corner zones always receive the overall lowest signal power, and the power to those corner zones is provided by the two APs which have non-zero gain to those corners. 
To mitigate the extreme effect of the scan loss approximation when the scan angle is larger than \ang{60} from broadside, we consider \SI{98}{\percent} signal coverage instead of \SI{100}{\percent}. 
In this way, the locations which are in endfire direction of the arrays are ignored. 
\SI{98}{\percent} signal coverage can be realized by ignoring the \SI{2}{\percent} locations which receive the overall lowest signal power when considering Constraints \eqref{eq:coherent} or \eqref{eq:non_coherent}.
Notably, in practice, there are always some multipath components that will help to cover the \SI{2}{\percent} locations.

\subsection{Power Saving Using the Optimized D-mMIMO}
Existing studies~\cite{dai2011comparative} have demonstrated that using the DAS reduces the distance between user locations and APs, and therefore, saves power.
Here we have extended the study to massive MIMO including the impact of the radiation pattern on the performance. 

\begin{figure}[!htb]
  \centering
  \includegraphics[width=0.8\linewidth]{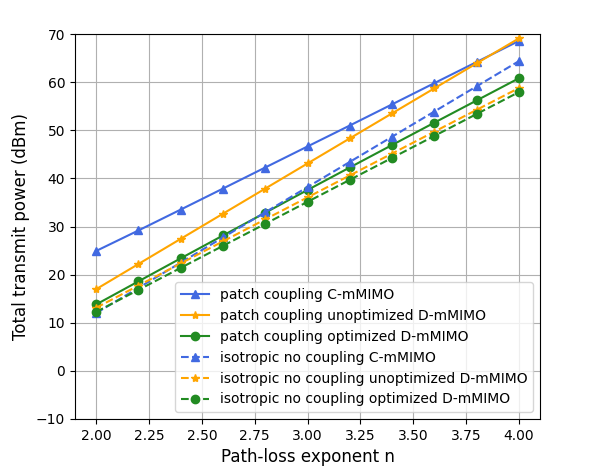}
  \caption{Transmit power comparison between the C-mMIMO and D-mMIMO considering different gain patterns and AP placements}%
  \label{fig:cas_das}
\end{figure}

Fig.~\ref{fig:cas_das} compares the total transmit power when using C-mMIMO, D-mMIMO with unoptimized AP placement (i.e., all the arrays are naively placed at the center of their edges), and D-mMIMO with optimal AP placement.
The considered region is a \SI{400}{\metre}$\times$\SI{400}{\metre} square.
The C-mMIMO has a single AP which has a 128 (i.e., 8$\times$16)-element uniform rectangular array and is deployed at the middle of an edge. 
Signals from different APs in the D-mMIMO system are non-coherently combined by using Eq.~\eqref{eq:non_coherent}.

Two types of embedded element gain patterns are considered in Fig.~\ref{fig:cas_das}, which represent two extreme cases regarding the directivity. 
`Isotropic no coupling' means the considered embedded gain pattern is non-directive, in other words, $\widetilde{G}_{\mathtt{E}_{\theta_{it}}}$ is fixed to one.
`Patch coupling' means considering the pattern of a patch antenna with the mutual coupling effect (Eq.\eqref{eq:element_gain}). The array gain pattern of each AP, in this case, has high directivity.

\begin{figure}
    \centering
    \subfloat[Isotropic no coupling, $n=2$.]{\includegraphics[width=0.22\textwidth]{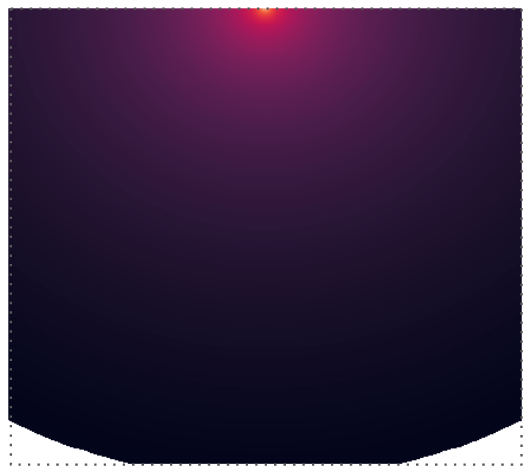}\label{fig:n2C-mMIMO}} 
    \subfloat[Patch coupling, $n=4$.]{\includegraphics[width=0.22\textwidth]{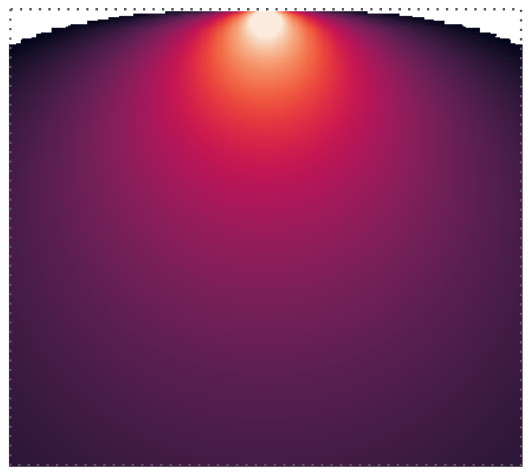}\label{fig:n4C-mMIMO}}
    \caption{Heatmaps of the received power for a single 128-element AP located at the top center. The white areas depict the locations where the received power is below the predefined threshold $P_{\mathtt{R}}$.
    When the radiation pattern is non-directive (a) the white areas are at the opposite side of the AP. When the radiation pattern is highly directive (b), the white areas are at the same side of the AP.
    This reveals that the benefit of using the D-mMIMO over the C-mMIMO is reducing the propagation distance in situation (a) and providing higher gain in situation (b). 
    }
\end{figure}

As illustrated in Fig.~\ref{fig:cas_das}, when considering the non-directive gain pattern, the power saving of the D-mMIMO over the C-mMIMO increases as the path-loss exponent increases. 
The reason is that when using C-mMIMO with the non-directive gain pattern, the locations that receive the minimum required signal power are on the opposite side of the AP even when the path-loss exponent is two (see Fig.~\ref{fig:n2C-mMIMO}). In other words, the long propagation distance between the AP and possible user locations is the barrier to reducing power here. 
Using the D-mMIMO, in this case, reduces the propagation distance, therefore allowing transmission at lower power, with the saving increasing as the path-loss exponent increases. 

When considering the highly directive gain pattern, the power saving of the D-mMIMO over the C-mMIMO decreases as the path-loss exponent increases.
This can be explained as when using the C-mMIMO considering the highly directive gain pattern, the particular locations are on the same side of the AP even if the path-loss exponent equals four (see Fig.\ref{fig:n4C-mMIMO}). In other words, the low gain around the endfire direction is the barrier for the power consumption here.
Using the D-mMIMO, in this case, provides higher gain to these locations.
However, this is at the cost of larger propagation distances between the APs and possible user locations. 

The results in Fig.~\ref{fig:cas_das} demonstrate that
\begin{enumerate*}[label={(\roman*)}]
  \item optimizing the AP placement is essential for improving the D-mMIMO's performance, the D-mMIMO with unoptimized AP placement may perform even worse than the C-mMIMO;
  \item the power saving by using the D-mMIMO compared to the C-mMIMO also depends on the array radiation patterns of the APs;
  \item when using patch arrays and considering mutual coupling, the power saving by using the D-mMIMO with optimized placement compared to the C-mMIMO is more than \SI{7}{\decibel} and up to \SI{11}{\decibel}.
\end{enumerate*}
 

\subsection{Impact of Different Factors on the D-mMIMO}
\begin{figure}[!htb]
  \centering
  \subfloat[
    The optimal AP placements in different scenarios, each placement is symmetrical regarding the region center and represented by the ratio between the x-coordinate of the top AP and edge length of the square region. The smaller value for the ratio, the closer the top AP to the left corner of the top edge.   
  ]{\includegraphics[width=0.4\textwidth]{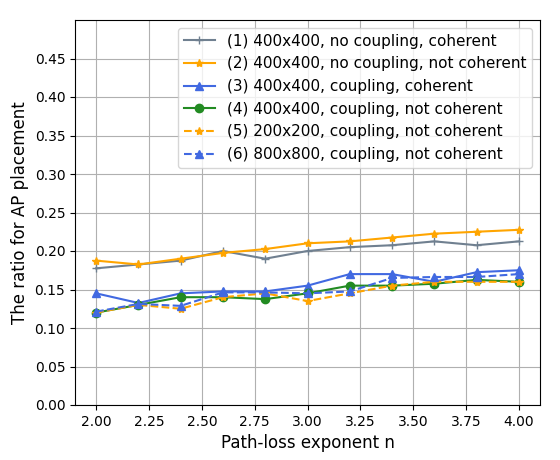}\label{fig:placement_impact}}\\\vspace{-10pt}
  \subfloat[The required total transmit power in different scenarios
  ]{\includegraphics[width=0.4\textwidth]{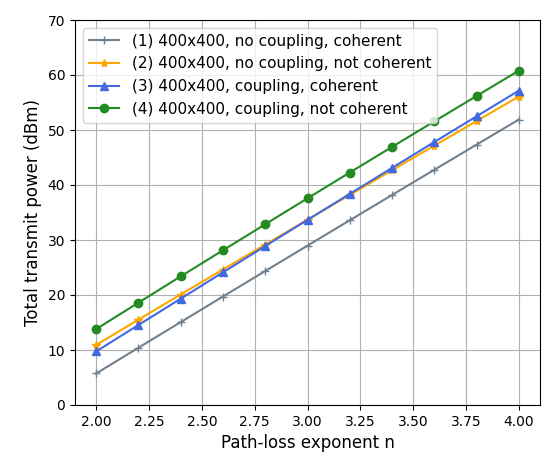}\label{fig:power_impact}}
  \caption{Impact of different factors on the optimal AP placement of the D-mMIMO, the factors include region size, mutual coupling, and coherence between the signals from different APs.}
  \label{fig:impact}
\end{figure}

This section investigates how different factors impact the transmit power and the optimal placement for the D-mMIMO that uses patch antennas. 
The factors include region size, mutual coupling between elements in each array, and coherence between the signals from different APs. 
The results for (1)-(6) scenarios are illustrated in Fig.~\ref{fig:impact}. 

The notation in Fig.~\ref{fig:impact}, for example, 
`no coupling' means the mutual coupling effect is not considered for the array gain (i.e., $\widetilde{G}_{\mathtt{E}_{\theta_{it}}} =  G_{\mathtt{E}_{\theta_{it}}}$). 
`Coherent' means the signals from different APs are coherent, using Eq.~\eqref{eq:coherent}. 
`200$\times$200' means the region is a \SI{200}{\metre}$\times$\SI{200}{\metre} square. 

As mentioned in Fig.~\ref{fig:placement}, each AP placement is symmetric regarding the center of the square region. 
Therefore, each placement can be represented by the ratio between the x-coordinate of the top AP and the edge length of the region. 
The smaller value for the ratio, the closer the top AP to the left corner of the top edge.   
Fig.~\ref{fig:placement_impact} illustrates the values for the ratio in different scenarios. 
The results demonstrate that the considered factors have little impact on the optimal placement except for the mutual coupling effect.

\begin{figure}[!htb]
  \centering
  \subfloat[With coupling]{\includegraphics[width=0.2\textwidth]{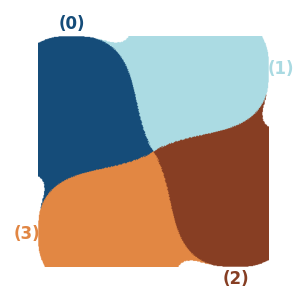}\label{fig:coupling}}
  \subfloat[No coupling]{\includegraphics[width=0.2\textwidth]{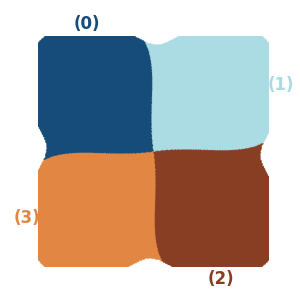}\label{fig:no_coupling}}
  \caption{The major signal contributor of each location considering the optimized D-mMIMO of two situations regarding the mutual coupling, (0)-(4) corresponds to the four APs. In the situation with mutual coupling, each AP needs to be the major signal contributor for much further away locations compared to the situation without mutual coupling.
  }
  \label{fig:mc}
\end{figure}

Fig.~\ref{fig:power_impact} shows the total transmit power required by the D-mMIMO in different scenarios. 
The results indicate that both mutual coupling and signal coherence have a large impact on the total transmit power. 
The power difference between scenarios (2) and (4) increases as the path-loss exponent increases. 
The reason for this is as follows. 
Fig.~\ref{fig:mc} illustrates the major signal contributors for different user locations when using the D-mMIMO with and without considering the mutual coupling effect. 
Different colors in Fig.~\ref{fig:mc} represent different APs. 
As illustrated in Fig.~\ref{fig:mc}, when the mutual coupling effect is considered, each AP has to provide signals to the locations which are much further away compared to the situation when the mutual coupling effect is not considered. 
This increases the propagation distance between the APs and the locations, therefore, being impacted more when the path-loss exponent increases. 
The power difference in scenarios (3) and (4) decreases as the path-loss exponent increases. 
The reason is that as the path-loss increases, the difference between the power of the signals from different APs increases, which reduces the power saving obtained by coherent combining of the signals. 
The power difference in scenarios (1) and (4) increases as the path-loss exponent increases.

\vspace{-10pt}

\FloatBarrier
\section{Conclusion}\label{sec:conclusion}
In this paper, we have studied rectangular topologies for the access point (AP) placement for the distributed massive MIMO (D-mMIMO).
Based on this topology, the AP placement is optimized to reduce the total transmit power. 
The impact of different factors including mutual coupling, region size, and signal coherence on the performance of the D-mMIMO is investigated. 
The results show that an optimized AP placement is essential for the performance of the D-mMIMO and the APs' radiation patterns have a high impact on the optimal AP placement.
With the optimized AP placement and using patch antennas, the D-mMIMO can save up to \SI{11}{\decibel} transmit power compared to the centralized antenna system with massive MIMO. 
For future research, it would be interesting to incorporate user interference in the problem model and to explore other promising topologies for the AP deployment. 

\section*{Acknowledgment}
We would like to thank Huawei for their support and NVIDIA for providing the GPU that greatly accelerated our simulations.
\bibliographystyle{IEEEtran}
\bibliography{literature} 

\end{document}